# Alternative modelling and inference methods for claim size distributions

Mathias Raschke, independent researcher/freelancer (www.mathiasraschke.de), mathiasraschke@t-online.de, Stolze-Schrey Str. 1, Wiesbaden, Germany

Abstract: The upper tail of a claim size distribution of a property line of business is frequently modelled by Pareto distribution. However, the upper tail does not need to be Pareto distributed, extraordinary shapes are possible. Here, the opportunities for the modelling of loss distributions are extended. The basic idea is the adjustment of a base distribution for their tails. The (generalised) Pareto distribution is used as base distribution for different reasons. The upper tail is in the focus and can be modelled well for special cases by a discrete mixture of the base distribution with a combination of the base distribution with an adapting distribution via the product of their survival functions. A kind of smoothed step is realised in this way in the original line function between logarithmic loss and logarithmic exceedance probability. The lower tail can also be adjusted. The new approaches offer the opportunity for stochastic interpretation and are applied to observed losses. For parameter estimation, a modification of the minimum Anderson Darling distance method is used. A new test is suggested to exclude that the observed upper tail is better modelled by a simple Pareto distribution. Q-Q plots are applied, and secondary results are also discussed.

## 1 Introduction

The modelling and estimation of the distribution of single claims of property lines of business is one of the main topics of actuarial science. The most important model for this issue is the Pareto distribution, which is frequently applied for the upper tail of the claim size distribution. The generalised Pareto distribution (GPD) is its extension. The application of the GPD to high extremes of Danish fire loss data by McNeil (1997) and the corresponding discussion by Resnick (1997) are well-known. The generalised Burr-gamma distribution was suggested as a complex model for the entire claims by Beirlant et al. (2000). Other approaches have kept the focus on the separate modelling of different parts of the distribution, called spliced distributions (Gian Paolo and Nino 2014, Finan 2017). Reynkensa et al. (2017) extended this approach on censored data.

A further opportunity is the discrete mixing of claim size distributions (Gian Paolo and Nino 2014, Finan 2017). A simple example of rail car damage (EmcienScan 2018) is shown in Figure 1 to illustrate the concept of a discrete mixture. The unknown cumulative distribution functions (CDF) are approximated by the empirical distribution function (EDF) in the example with;

$$\hat{F}(x_i) = \frac{i}{n+1}. \tag{1}$$

for observation $x_i$ of the ordered sample of size $n$ with observations $x_1 \leq \cdots \leq x_i \leq \cdots \leq x_n$. The distribution for all damage in Figure 1 is obviously the discrete mixture

$$F(x) = (1-P)F_1(x) + PF_2(x), 0 \leq P \leq 1. \qquad (2)$$

In the example, CDF $F_1$ applies to all damage which does not have the departure Tennessee (TN), $F_2$ applies to damage with departure TN. The reason for this extraordinary mixture is not researched here.

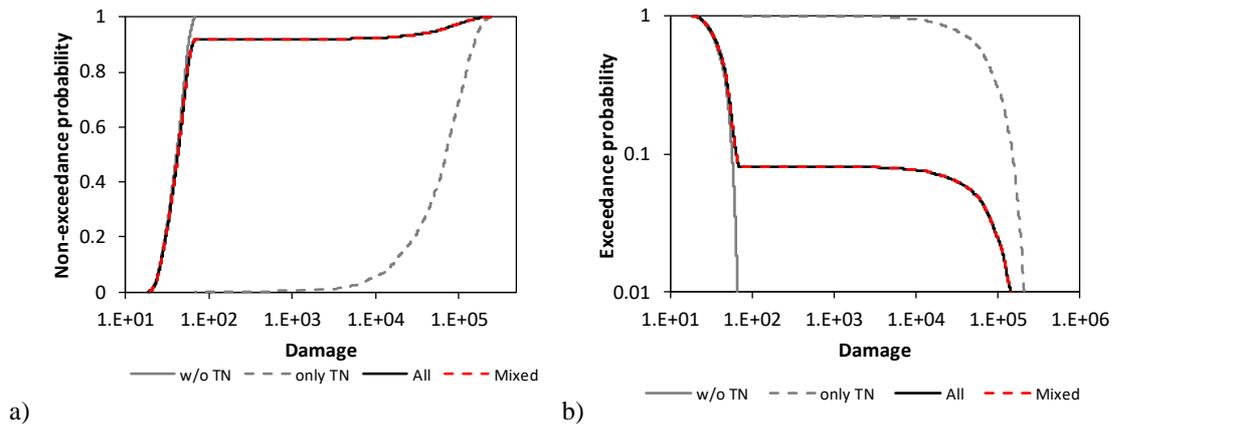

a) b)
Figure 1: Mixing of EDF for the train car damage with $\hat{P} = 0.081$ for the share of TN according to the sample: a) CDF and b) survival function.

Gian Paolo and Nino (2014) applied a discrete mixture to the Danish fire loss data. Additionally, a complex dynamic mixture was suggested by Frigessi et al. (2002) for the Danish fire loss data, and Bakar et al. (2015) call their spliced distribution composite model for this data since they compose a Weibull distribution with a second one. The advantage of their approach is that the threshold is estimated during the inference procedure. The optimal threshold selection is an issue in classical modelling of claim sizes with a Pareto tail. Scarrott and McDonald (2012) present a review about the different approaches to dealing with this issue. There are also new opportunities in extreme value statistics such as for the estimation of the extreme value index (Hüsler et al. 2016). Beirlant et al. (2016 and 2017) presented new estimation methods for a truncated tail. However, a truncated tail can be a mixture of cut tails as shown by Raschke (2014) for earthquake magnitudes. The resulting distribution is formulated by the product of survival functions. The product of survival functions has already been used by Meerschaert et al. (2012) for exponentially tempered power law distributions.

The examples of models mentioned are diverse. The patterns of observed claim sizes are more diverse. Therefore, a new approach is introduced in section 2 which extends the opportunities for claim size modelling. It combines the classical modelling via Pareto distribution or GPD with the approach of Raschke (2014) and Meerschaert et al. (2012) with the opportunity of discrete mixing. The tails of the base distribution are adjusted by these approaches. Attention is paid to the generating of a random mechanism to provide the opportunity for a reasonable stochastic interpretation of the complex claim generating process. The latter includes unknown physical, social, and economic elements. The corresponding extreme value index of the upper tail is also discussed.

The special inference methods are applied for the new approach and are explained in section 3. The minimum Anderson Darling distance (MAD) method of Boss (1982) is modified and is a variation of the maximum likelihood method (Raschke 2017). The MAD has already been applied to claim/loss data by Clemente et al. (2012), Skřivánková, and Juhás (2012) and Gian Paolo and Nino (2014). The computational procedure is briefly explained, and a new test is suggested to exclude the possibility that a simple separate Pareto distribution would be a reasonable alternative for the very upper tail.

In section 4, the MAD is applied to Norwegian fire loss data to demonstrate its advantages and performance. Then, in this section, the new model approach is used in combination with the explained inference methods to AON Re claim data, the Danish fire loss data and Asian claim data. At the end of the paper, the main and secondary results are discussed followed by a conclusion.

It is underlined that, in all, cases independence between the positive real valued random variables and between the realizations of each single random variable is assumed in this paper. Furthermore, the analysed random variables are damage, losses or claims. The terms are synonymous in most cases. When not, the differences are explained.

All analysed data are provided by the R-cran package 'CASdatasets' by Dutang (2016), being developed for the book, edited by Charpentier (2014). The corresponding R-manual also describes the original data source.

## 2   Basics of the modelling

### 2.1   The model idea

The claim generating process is complex, with unknown physical, social and economic elements. Therefore, claim size is also influenced by the claim size management of an insurance policy. Medium size claims are liable to a 'normal' regime, which is considered by the base distribution with CDF $F_{base}$ and its survival function $\bar{F}_{base}$. Very high damage, higher than the upper threshold $x_{upper}$, is reviewed more in detail. The more rigorous claims management results in lower claim sizes than the 'normal' claim management. This can be considered by the minimum principle of the sample minimum being used here for two different random variables with;

$$X = Min\{Y, W\}. \tag{3}$$

In this stochastic mechanism, $Y$ represents the 'normal' claim size management and $W$ represents the influence of the more rigorous claims management. It can be interpreted as an independent random right cut-off point of the distribution of $Y$ (Raschke 2014). The survival function of the resulting continuous random variable $X$ (final claim size) is the product of the survival functions of $Y$ and $W$. These are the survival functions of the base distribution for $Y$ and the (upper) adjusting distribution for $W$ in the upper tail adjustment;

$$\bar{F}_{upper}(x) = \bar{F}_{base}(x)\bar{F}_{adj,u}(x). \tag{4}$$

Meerschaert et al. (2012) used this approach for the tempered Pareto distribution, and Raschke (2014) used this approach for earthquake magnitude distributions. The stochastic mechanism (3) was already explained in the latter references. The interpretation for claim size distributions is the contribution of this current paper. Additionally, the approach is extended here. The special management for high claims could be activated only for a certain line of business or could fail and is only (correctly) applied with a certain probability - the transition probability, $P_{upper}$. The base distribution $F_{base}$ would also apply for the share $1 - P_{upper}$ of the high claims. The final adjusted upper tail is a discrete mixture of the unadjusted survival function of the base distribution and the product of survival functions of the base distribution and adjusting distribution;

$$\bar{F}_{upper}(x) = \bar{F}_{base}(x)\bar{F}_{adj,u}(x)P_{upper} + \bar{F}_{base}(1 - P_{upper}). \tag{5}$$

The medium part of the base distribution should not be modified; the condition,

$$F_{upper}(x) = F_{base}(x), if\ x \leq x_{upper} \tag{6}$$

should (approximately) hold, which (approximately) means that;

$$\bar{F}_{adj,u}(x) = 1, if\ x \leq x_{upper} \tag{7}$$

Meerschaert et al. (2012) and Raschke (2014) have suggested an adjustment only for the upper tail. Here, the lower tail is also adjusted. It can be done independently of the upper tail adjustment since conditions (6-7) are held. For the lower tail adjustment, a maximum principle with respect to the sample maximum;

$$X = Max\{Y, W\} \tag{8}$$

is assumed for the management of small claims. The 'normal' regime would result in a random variable $Y$. The special considerations for the reduction of the administrative overhead are represented by random variable $W$. It is highlighted that this is a rough interpretation. However, the lower tail of the claim size distribution can be adjusted by the product of $F_{base}$ and the adjusting distribution $F_{adj,l}$ according to (8) and the extreme value theory (cf. Beirlant et al. 2004)

$$F_{lower}(x) = F_{base}(x)F_{adj,l}(x). \tag{9}$$

A mixing as for the upper tail (5) is also possible for the lower tail but is not considered further in this research. The condition;

$$F_{lower}(x) = F_{base}(x), if\ x \geq x_{lower} \tag{10}$$

should (approximately) hold, and is similar to the condition in (6), and implies (approximately) that;

$$F_{adj,l}(x) = 1, if\ x \geq x_{upper}. \tag{11}$$

In all cases is $x_{upper} > x_{lower}$ or better $x_{upper} >> x_{lower}$. The principle effect of the adjustments is shown in Figure 2. All variants of different transition probabilities result in the same adjustment of the lower tail since this is determined only by the adjusting distribution $F_{adj,l}(x)$. The upper tail of the adjusted claim size distribution depends on $P_{upper}$. If $P_{upper}<1$ and the base distribution is a Pareto one, then the graph of the adjusted upper tails of the survival function runs asymptotically parallel to the base distribution in the space of logarithmic exceedance probability and logarithmic loss/claim. This can be seen well in Figure 2c. Therein, the thresholds of the adjustments are marked.

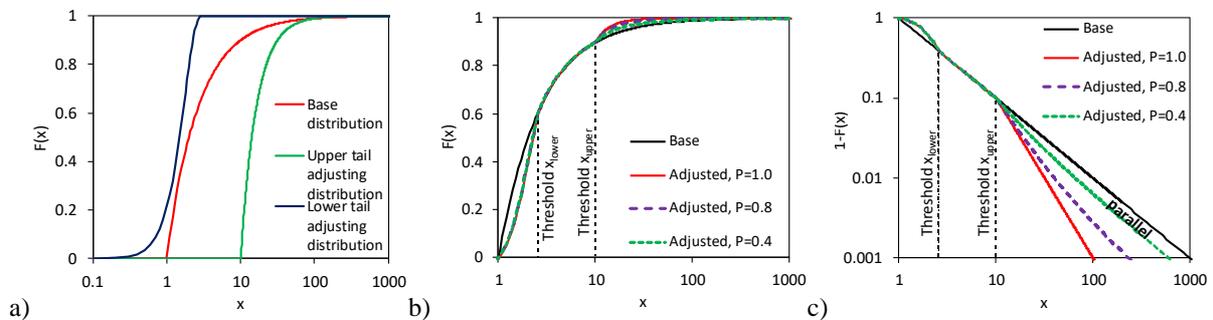

Figure 2: Fictitious example of the adjustment: a) CDFs considered distribution models, b) resulting adjusted CDF for different transition probabilities $P_{upper}$, c) corresponding survival functions.

The well-known Pareto and/or the generalised Pareto distributions are the 'natural' models for the base distribution since these are frequently used as claim size distributions. These are formulated in the next sub-section. A special phenomenon of the Pareto distribution is also discussed which influences the consideration of inflation in processing the observations. Two further subsections deal with approximations and alternative generating mechanism for the adjustments of the lower and upper tail.

## 2.2 The Pareto distribution as base distribution and the inflation issue

The Pareto distribution has the survival function with extreme value index (tail index) $\gamma$ or Pareto-$\alpha$

$$\bar{F}(x) = 1 - F(x) = \left(\frac{\sigma}{x}\right)^\alpha = \left(\frac{x}{\sigma}\right)^{-\frac{1}{\gamma}}, x \geq \sigma, \alpha = 1/\gamma, \alpha > 0 \qquad (12)$$

and the generalised Pareto distribution (GPD; Beirlant et al. 2004) has survival function

$$\bar{F}(x) = \begin{cases} \left(1 + \gamma \frac{x}{\sigma}\right)^{-\frac{1}{\gamma}}, x \geq 0, \sigma > 0, if\ \gamma \neq 0 \\ exp\left(-\frac{x}{\sigma}\right), x \geq 0, \sigma > 0, if\ \gamma = 0 \end{cases}, \qquad (13)$$

and are used here as base distribution for the following reasons. Firstly, the conditional tail distribution (distribution of peaks over threshold) of many distributions converges to a (generalised) Pareto distribution with the threshold converging to the right (infinite) endpoint (cf. Beirlant et al. 2004). Secondly, the Pareto distribution or the generalised Pareto distribution with $\gamma > 0$ (Fréchet domain of attraction) is frequently successfully applied to insurance claims. Thirdly, the shape of empirical loss distributions often forms a straight line with logarithmic loss and logarithmic exceedance probability, at least in a certain range (cf. examples in section 4).

Furthermore, the Pareto distribution is also the conditional distribution of a realisation of a Poisson process at the line of real numbers with intensity $\lambda(x) \sim x^{-\alpha-1}$. The latter is part of the construction of max-stable random fields in Schlather's (2002) 2nd Theorem (with $\alpha = 1$). The occurrence of claims with a certain size is a point process, even though only the claim size distribution is considered in the current paper. It is a conditional distribution with the condition that a claim event occurred.

The theorem of Schlather (2002) also implies an interesting item. If $X$ are the points of a realisation of a point process and $Y$ is a random variable with $Y > 0$ and a finite expectation, then a realisation of a further point process is generated with points $W = YX$. The intensity of the new point process also has the intensity $\lambda(x) \sim x^{-\alpha-1}$. This fact means that the claim observations do not need to be inflated for the compensation of inflation effect in the estimation of $\alpha$ and $\gamma$ of (12), in contrast to the estimation of claim frequency. A fictitious example illustrates this principle. The observed claims are a realisation of the Poisson process with density $\lambda(x) = 1{,}05^{k-1} x^{-\alpha-1}$ for the $k$th year from 1 to 10 (Figure 3a) with $\alpha = 1$. Factor 1.05 means the inflation is 5% a year. The underlying conditional excess distributions are obviously the same, only the loss frequency is different. For the eleventh year, the correct conditional distribution can be formulated for an excess with a stationary threshold of 0.01 by the empirical distribution function of the un-inflated observations. In contrast, the stationary threshold and the inflated claims act like thinning and the excess distribution for the eleventh year is not correct (Figure 3b). However, the tail behaviours of inflated and not-inflated claims are the same; the empirical survival functions of Figure 3c are approximately parallel in the range of higher losses in the log-log plot. The inflated claims/losses would need a further threshold larger than 0.01 for the estimation of the claim size distribution with respect to its upper tail. A higher threshold implies a smaller sample size, a loss of information and a lower accuracy in the estimation of the extreme value index $\gamma$ and Pareto-$\alpha$ in (12-13). This secondary result is considered in a real-world example in section 4.1. The issue of the inflation of the deductibles of insurance policy is an issue which is not considered in this discussion in the current paper.

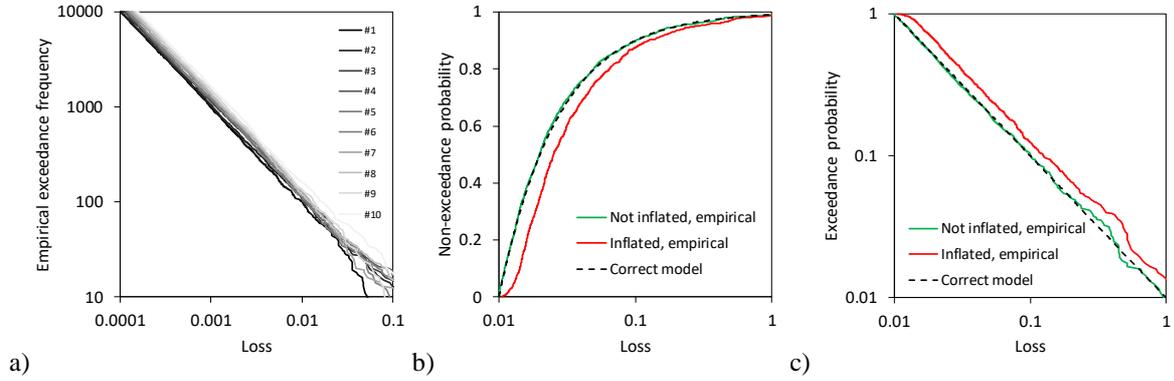

Figure 3: Example of claims as point processes and corresponding conditional distribution: a) aggregated occurrence functions (exceedance frequency) of the point processes per year 1 to 10, b) CDF for claims larger than 0.01 (inflated to eleventh year with 5% inflation a year), c) corresponding survival functions.

## 2.3   Approximations of the upper tail adjustment

Of course, the constructions of section 2.1 are simple and real life can be more complex. It might also be the case that an approximation could replace the suggested construction. The following example illustrates this. For the suggested model of section 2.1, the Pareto distribution (12) is the base distribution with extreme value index $\gamma = 1/\alpha = 1$ and scale $\sigma = 1$. The adjusting distribution is a shifted Weibull distribution with $x_{upper} = 3$, $\beta_{adj,u} = 2$ and $\sigma_{adj,u} = 25$ and survival function

$$\bar{F}(x) = exp\left(-\left(\frac{x-x_{upper}}{\sigma_{adj,u}}\right)^{\beta_{adj,u}}\right), x \geq x_{upper}, \sigma_{adj,u} > 0, \beta_{adj,u} > 0. \tag{14}$$

The base and adjusted distributions are shown in Figure 4 for transition probability $P_{upper} = 0.5$ and $P_{upper} = 1.0$. An approximation is also shown for the model with $P_{upper} = $ 0.5; it is a spliced distribution (Gian Paolo and Nino 2004, Finan 2017), which uses a kind of cascading or stepped Pareto distribution with breaking points;

$$\bar{F}(x) = \begin{cases} \left(\frac{x}{\sigma_1}\right)^{-\alpha_1}, & \text{if } \sigma_1 \leq x \leq \sigma_2 \\ \left(\frac{\sigma_2}{\sigma_1}\right)^{-\alpha_1}\left(\frac{x}{\sigma_2}\right)^{-\alpha_2}, & \text{if } \sigma_2 \leq x \leq \sigma_3 \\ \left(\frac{\sigma_2}{\sigma_1}\right)^{-\alpha_1}\left(\frac{\sigma_3}{\sigma_2}\right)^{-\alpha_2}\left(\frac{x}{\sigma_3}\right)^{-\alpha_1}, & \text{if } \sigma_3 \leq x \end{cases} \quad (15)$$

with parameters $\alpha_1 = 1$, $\alpha_2 = 1.42$, $\sigma_1 = 1$, $\sigma_2 = 11$ and $\sigma_3 = 52$. The PDF of the approximating survival function (15) is simply a function with three segments; each of them is determined by scaled PDF of a Pareto distribution. The extreme value index of the first segment and the last segment are equal to realize a similar survival function as the approximated model with adjusted upper tail. Details of the extreme value index are also discussed in the next section. Apart from this detail, the approximation performs well and the number of parameters of both models are equal. The question remains, which model reflects the claim generation process of the real world better?

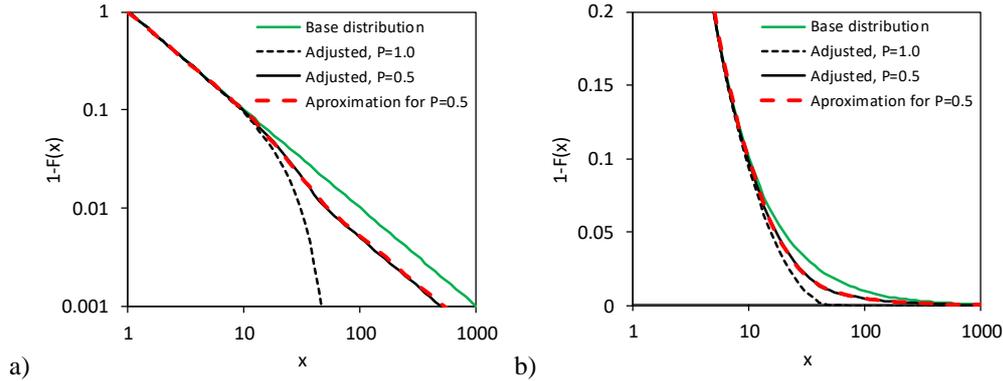

Figure 4: Example of a distribution model according to (9) and its approximation by (15): a) and b) shows the survival functions with different presentation of a scale.

## 2.4 Asymptotic behaviour of the upper tail adjustment

The main characteristic of the new approach is the possibility that the very upper tail has the same behaviour as the base distribution of the 'normal' claims. The transition range between is influenced by $\bar{F}_{adj,u}$ (8<x<50 in Figure 4) and acts like a smoothed step or smoothed shift of the original linear graph of function between logarithmic loss and logarithmic exceedance probability. The asymptotic distance between the upper tails of base and adjusted distribution is controlled by important transition probability $P_{upper}$. Accordingly (5), it is asymptotically (cf. Figure 2c and Figure 4a)

$$P_{upper} = \lim_{x \to \infty}\left(1 - \frac{\bar{F}_{upper}(x)}{\bar{F}_{base}(x)}\right). \quad (16)$$

The model is formulated for the instance where there is no finite right end point for $\bar{F}_{base}(x)$ and $\bar{F}_{adj,u}(x)$. It would not be difficult to consider such, but the infinite case is more important in practice, as far as the author is aware.

If $P_{upper} < 1$ then $\bar{F}_{upper}(x)$ has the same extreme value index (tail index) as $\bar{F}_{base}(x)$ since (16) applies. Otherwise, with $P_{upper} = 1$, the new extreme value index $\gamma_{upper}$ is;

$$\gamma_{upper} = \frac{1}{1/\gamma_{base} + 1/\gamma_{adj,u}}, \qquad (17)$$

with extreme value index $\gamma_{base}$ of the upper tail of the base distribution and extreme value index $\gamma_{adj,u}$ of the upper tail of the adjusting distribution. This can be easily derived by the product of the survival functions of two Pareto distributions with;

$$\alpha_{upper} = \alpha_{base} + \alpha_{adj,u}. \qquad (18)$$

Furthermore, if the adjusting distribution is of the Gumbel domain of attraction (Beirlant et al. 2004) with $\gamma_{adj,u}=0$, then $\gamma_{upper}=0$. This is the simple result for a limit value analysis of (18) with $\alpha_{adj,u} \to \infty$ and the relation $\gamma = 1/\alpha$ in (17).

## 2.5 *A note about the claim generating process of the lower tail*

In section 2.3, the issue of approximation of one model by another was discussed for the upper tail. In the current section, such an issue is discussed for the lower tail. This is done for the logarithm of losses/claims for reason of simplification. It is well known that a Pareto distribution is transformed in this way to an exponential distribution. The latter has the CDF;

$$F(x) = 1 - exp(-x/\sigma). \qquad (19)$$

As explained in section 2.1, the lower tail of the claim size distribution can be adjusted via the maximum principle of (8) which leads to (9) because there is the tendency of offering the policy holder slightly more than the actual claim size estimation to avoid administrative costs. A further mechanism of the real world is reasonable. Let us assume that in some cases the policy holder does not claim the replacement of the loss from the insurer since his/her own administrative costs and efforts are too high. In this instance, there is an actual difference between loss distribution and claim size distribution. The latter would be a thinned version of the first. The thinning probability depends on the loss size and is formulated with function $G(x)$. The thinning function should be a decreasing function and fulfil the condition $0 \leq G(x) \leq 1$. Therefore, a survival function of a distribution model is an appropriate model.

The formulation for the thinned distribution includes $G(x)$ and the probability density function $f(x)$ of the original loss distribution (here for lower threshold 0);

$$F_{thinned}(x) = \frac{\int_0^x f(y)G(y)dy}{\int_0^\infty f(y)G(y)dy} \quad . \tag{20}$$

A threshold $x_{upper}$ could be considered in the formulation if $G(x_{upper}) = 0$.

Let the thinning probability function be an exponential one;

$$G(x) = exp(-x/\sigma_t). \tag{21}$$

If the distribution of the logarithmic losses is exponential (19), then the thinned distribution – the claim size distribution is;

$$F_{thinned}(x) = 1 - \frac{(\sigma+\sigma_t)exp\left(-\frac{x}{\sigma}\right) - \sigma_t exp\left(-\frac{x(\sigma+\sigma_t)}{\sigma\sigma_t}\right)}{\sigma}. \tag{22}$$

As fictitious example, distributions with thinning and maximum principle are compared in Figure 5. The original, un-thinned base distribution is exponential according to (19) with scale parameter $\sigma = 1$. The thinning probability function (21) also has a scale parameter $\sigma_t = 1$. And the adjusting distribution of (9) is also an exponential distribution (19) with scale parameter $\sigma = 1$ ($x_{lower} = \infty$). Both, thinning and the maximum principle results in an equal distribution of the final, observed claims.

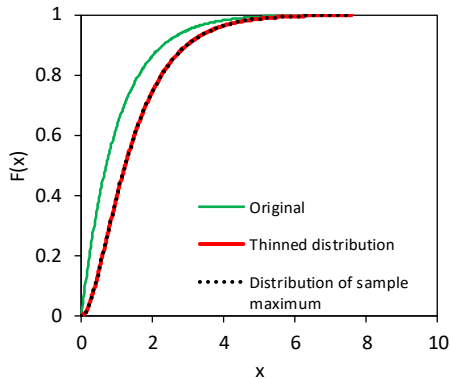

Figure 5: Example of claim size distribution being generated by thinning or maximum principle (scale $x$ represents logarithmic losses/claims).

The fictitious example demonstrates that the claim generating process of the real world cannot be identified only with the sample of claims. Additionally, the adjustment (9) can be used also as an approximation for completely different claim generating processes of the real world. An interesting detail – the differences between the original and thinned distribution in Figure 5 and the differences between the not-inflated and inflated losses of Figure 3b follow the same pattern.

## 3  Notes on the statistical inference

### *3.1  The Minimum Anderson Darling distance method*

As aforementioned, minimum distance estimation methods are already applied in actuarial science. Wolfowitz (1957) and Parr and Schucany (1980) have developed and discussed this family of methods, and Drossos and Philippou (1980) has shown that these methods are invariant. Here, the Minimum Anderson Darling distance (MAD) method (Boos 1982) is used and implies the minimisation of the Anderson Darling test statistic of Anderson and Darling (1954; c.f. Stephens 1986) for the ordered sample $x_1 \leq x_2 \leq \cdots \leq x_i \leq \cdots \leq x_n$ for a parameter vector $\boldsymbol{\theta}$

$$A(\boldsymbol{\theta}) = -n - \sum_{i=1}^{n}(2i-1)\ln(F(x_i; \boldsymbol{\theta})) + (2(n-i)+1)\ln(1 - F(x_i; \boldsymbol{\theta})). \quad (23)$$

The corresponding estimator is;

$$A(\hat{\boldsymbol{\theta}}) = \min_{\boldsymbol{\theta} \in \Theta} A(\boldsymbol{\theta}). \quad (24)$$

The MAD has already been used in actuarial science by Clemente et al. (2012), Skřivánková, and Juhás (2012) and Gian Paolo and Nino (2014). Raschke (2017) has shown that this estimator is a variant of maximum likelihood estimators. The corresponding mixed likelihood functions of a Bernoulli distribution is;

$$l_B(\boldsymbol{\theta}) = \frac{1}{n}\sum_{i=1}^{n}(i - 0.5)\ln(F(x_i; \boldsymbol{\theta})) + (n - i + 0.5)\ln(1 - F(x_i; \boldsymbol{\theta})), \quad (25)$$

and the estimator is;

$$l_B(\hat{\boldsymbol{\theta}}) = \max_{\boldsymbol{\theta} \in \Theta} l_B(\boldsymbol{\theta}). \quad (26)$$

One advantage of this estimation method is that the consideration of a part of the ordered sample (sum in (25) for a subrange of [1,n]) does not lead to an (asymptotic) bias but only to a higher estimation error if the distribution model is correct. The performance of MAD is good for a GPD (Raschke 2017). Additionally, a weighting by the order position can be considered. The focus can be on the upper tail in this way. Such a weighting is not possible for a conventional maximum likelihood estimation since the weighting corresponds to a weighted empirical distribution function. This has the same consequence as the underlying distribution which is changed since the differences between empirical distribution function and actual distribution function are asymptotically negligible according to the well-known Glivenko–Cantelli theorem. In addition, the MAD method (25-26) is modified here to;

$$l_B(\boldsymbol{\theta}) = \frac{1}{n}\sum_{i=1}^{n} w_i(i\ln(F(x_i; \boldsymbol{\theta})) + (n - i + 1)\ln(1 - F(x_i; \boldsymbol{\theta}))). \quad (27)$$

This modification implies that every summand has its maximum with $F(x_i; \theta) = \frac{i}{n+1}$; this is the expectation $E[F(x_i; \theta)]$ according to order statistics (cf. Balakrishnan and Clifford Cohen 1991, Johnson et al. 1995). The first suggested weighting is a normalisation of each summand at this expectation with;

$$w_i = \frac{1}{i \ln(i/(n+1)) + (n-i+1)\ln(1-i/(n+1))}. \tag{28}$$

A further weighting is also considered in section 4.1 with preference of the largest observation;

$$w_i = \frac{\sqrt{i}}{i \ln(i/(n+1)) + (n-i+1)\ln(1-i/(n+1))}. \tag{29}$$

In summary, the MAD is very flexible since it needs only sub-samples of segments of a distribution, the performance is good, and it is a variation of the ML method and a weighting can be easily considered. In addition, the author is familiar with the method. Therefore, this method is applied. The performance is also demonstrated by the example in section 4.1.

The estimation error can be computed with a resampling method (Efron 1979, Davidson and Hinkley1997). The Bootstrap approach is used here. The entire sample is resampled here without special consideration.

### 3.2 Normalised spaces as an opportunity to estimate Pareto-α

If the base distribution is modelled by a Pareto distribution, then it is a further opportunity to estimate Pareto-α. The Hill estimator (Hill 1975) implies an estimation of the expectation of an exponential distributed random variable since the logarithmic losses are exponentially distributed. The normalised spaces of the ordered sample of an exponentially distributed random variable also results in an exponentially distributed random variable with the same expectation according to Sukhatme (1936). This offers the opportunity of estimating the parameter using only observations within a defined range of the loss/claim scale. Raschke (2014) applied this approach on the distribution of earthquake magnitudes. However, this approach is only mentioned in passing and is not used in the following examples.

### 3.3 Computational steps

The parameters of the base distribution are first estimated since this is the basis of the modelling. For this, the MAD method is applied to all observations in the range between $x_{lower}$ and $x_{upper}$. The thresholds $x_{lower}$ and $x_{upper}$ must be selected visually, as for spliced models. It is possible, that the adjusting distributions fulfil condition (7,11) approximately without explicit consideration of $x_{lower}$ or $x_{upper}$ in the parametrization of the adjusting distribution. The thresholds $x_{lower}$ and $x_{upper}$ must be defined nevertheless since they determine the sub-samples for the parameter estimation. In a second and third step, the parameters of the adopting distribution (and the mixing) are estimated for the upper and lower tails. Observations smaller than $x_{lower}$ are considered for the lower tail adjustment and observations larger than $x_{upper}$ for the upper tail adjustment. The conditions (6,7,10,11) have to be fulfilled (approximately). Simple numerical optimisations can be, and are, used in the parameter estimation.

*3.4 Goodness-of-fit*

The fit cannot be quantified in the meaning of an exact Goodness-of-fit test. Simple visual checks are suggested. A Q-Q plot is applied here as already used for claim size data by Beirlant et al. (2004) and Gian Paolo and Nino (2014). The Q-Q plot is applied for the original margins and/or for standard normal margins and standard Fréchet margins.

However, a visual check is subjective; a quantitative check is desired. The comparison of different Pareto-α being estimated for different ranges of the loss scale by the normalised spaces (section 3.2) would be an opportunity if there were enough observations of the upper tail. The estimations of Pareto-α should not be stable in the case of adjusted tails. However, too many observations are needed for this idea. A further opportunity is to test the goodness-of fit of a separate Pareto-Model for the upper tail. This could be done via a goodness-of-fit test for exponential distributions (e.g. by Stephens 1984). However, this test of little use if the sample size is small, as is so often the case. Therefore, a new test is suggested for hypothesis $H_0$ of a Pareto distribution for the $k$ largest observations. The corresponding threshold $x_{n-1-k}$ of the ordered sample should be in the transition range. One of the biggest differences between an (unknown) adjusted distribution and a (perhaps) falsely assumed Pareto tail (12) is that the actual graph of logarithmic loss to logarithmic exceedance probability is not a straight line but a concave line in the transition range. This can be tested by the largest number $m$ of observations in series of consecutive tail observations $x_i$ of the ordered sub-sample $x_1 \leq x_2 \leq \cdots \leq x_i \leq \cdots \leq x_k$, which fulfil the condition $\hat{F}(x_i) > F_{Pareto}(x_i)$. In detail, at first the numbers $l_i$ are computed;

$$l_i = \begin{cases} l_{i-1} + 1 & \text{if } \hat{F}(x_i) > F_{Pareto}(x_i), \ l_0 = 0 \\ 0 & \text{if } \hat{F}(x_i) \leq F_{Pareto}(x_i) \end{cases} \tag{30}$$

The test statistic is computed with;

$$m = \max_i l_i. \tag{31}$$

The Pareto-$\alpha$ is estimated by the Hill-Estimator; scale parameter $\sigma$ in (12) is $x_{n-1-k}$ of the entire sample. The probability that $m$ equals or exceeds the current value can be computed for the hypothesis $H_0$ of a Pareto distribution by Monte Carlo simulations, here with 10,000 repetitions. The computed probability should be larger than the defined significance level (here 5%) for the acceptance of the $H_0$. The Hill estimator (Hill 1975) is also used in these simulations.

A fictitious example illustrates the suggested test procedure. The actual adjusted tail is presented in Figure 6a. The actual threshold for the tail adjustment is $x_{upper} = 10$. The sample is also depicted. In Figure 6b, the estimated base distribution is shown. Pareto-$\alpha$ was estimated by using the normalised spacing (section 3.2); all observations smaller than 8.22 were considered. The tail sample and the separate tail model are shown in Figure 6c. The threshold is $x_{n-1-k} = 11.934$ or $k = 61$. The green points in Figure 6c represent the observations which fulfil the condition $\hat{F}(x_i) > F_{Pareto}(x_i)$. The observations in the longest series of consecutive tail observations with $\hat{F}(x_i) > F_{Pareto}(x_i)$ are marked in black. It seems that there is only one series. However, in a section enlarged in Figure 6d; there are two series with observations $\hat{F}(x_i) > F_{Pareto}(x_i)$. The first includes only one observation. The test statistic is $m = 43$. The probability that a correct Pareto model with $k = 61$ results in such a value or higher is only 3.4%. The hypothesis $H_0$ of a Pareto tail is rejected for 5% significance level.

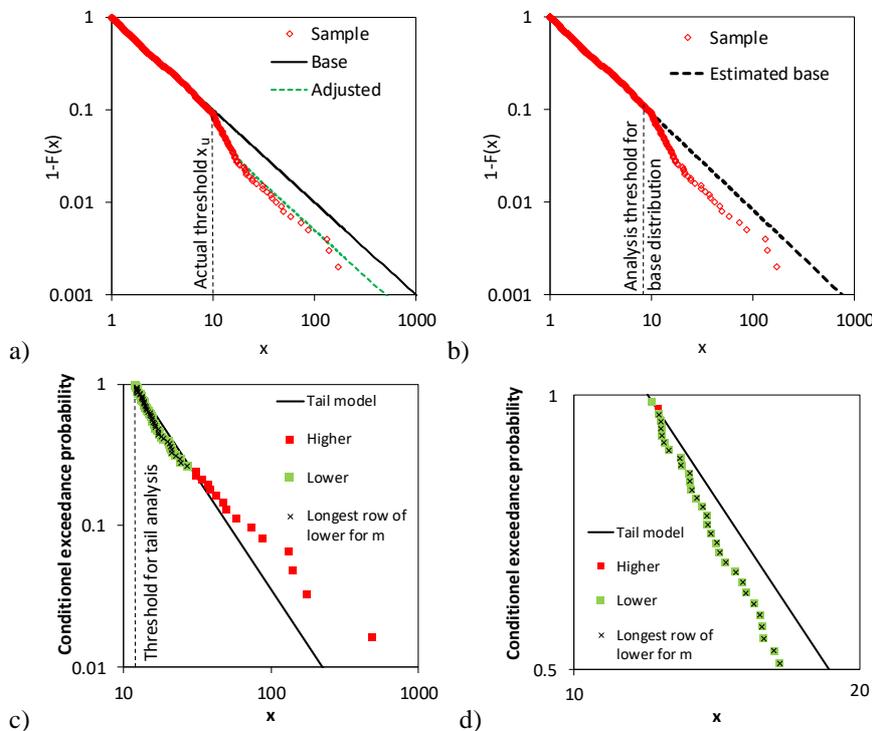

Figure 6: Example for the quantitative test for Pareto tail with focus on non-concave shape: a) actual distribution and sample, b) estimated base distribution and sample, c) tail analysis with Pareto model and test observations, and d) section of c).

## 4 Examples

### 4.1 Norwegian fire loss data

The performance of the MAD method is demonstrated in the first example. Therein, all observations of the Norwegian fire loss data ($n = 9181$, from 1972 to 1992, unit 1000 Krone) are considered. Beirlant et al. (2004) have already analysed the observations of 1976, and Beirlant et al. (2001) have built a model for the entire data. However, it is not clear if Beirlant et al. (2001) used exactly the same data which have been analysed here (raw loss data of Dutang 2016).

The minimum loss is 500 and was observed 161 times. Therefore, the application of an excess threshold of 499 is reasonable. The GPD is fitted with the ML method and variants of the MAD method. All alternatives fit the observations well according to Table 1 and Figure 7.

Table 1: Estimated parameters for GPD for the Norwegian fire loss data (the versions of the MAD method are distinguished by roman numbers)

| Method | Weighting | Sample size $n$ | $\hat{\gamma}$ | $\hat{\sigma}$ |
|---|---|---|---|---|
| ML | No | 9181 | 0.649±0.017 | 599.96±11.53 |
| MAD I | according to (28) | 9181 | 0.667±0.018 | 589.90±12.32 |
| MAD II | according to (29) | 9181 | 0.662±0.017 | 592.83±11.67 |
| MAD III | according to (28) | 5000, $i$=4182 to 9181 in (24) | 0.680±0.019 | 574.65±12.42 |

The different weighting of the MAD method does not have a significant influence and the MAD fit for the 5000 largest observations does lead to very different estimations.

Beirlant et al. (2001) applied the Burr distribution with more than two parameters. Their Q-Q plot does not indicate a better fit than the current Figure 7c and b.

Dutang (2016) also provide loss data scaled at the level of the year 2012. The two empirical survival functions have a similar difference as the actual function and the function for the corrected observations in the fictitious example (Figure 3c). This supports the secondary result of section 2.2. In addition, the advantage of the analysis of raw un-inflated data can be demonstrated. The Hill estimator (Hill 1975) is applied for estimation of the extreme value index of raw data with a threshold of 7,000, which results in $\hat{\gamma} = 0.684 \pm 0.034$. The inflated data for 2012 can be analysed for a corresponding threshold of 45,030.3 (maximum scaling to compensate inflation is 6.432) and leads to $\hat{\gamma} = 0.677 \pm 0.067$. The point estimation is very similar; however, the standard errors strongly differ because of different sample sizes.

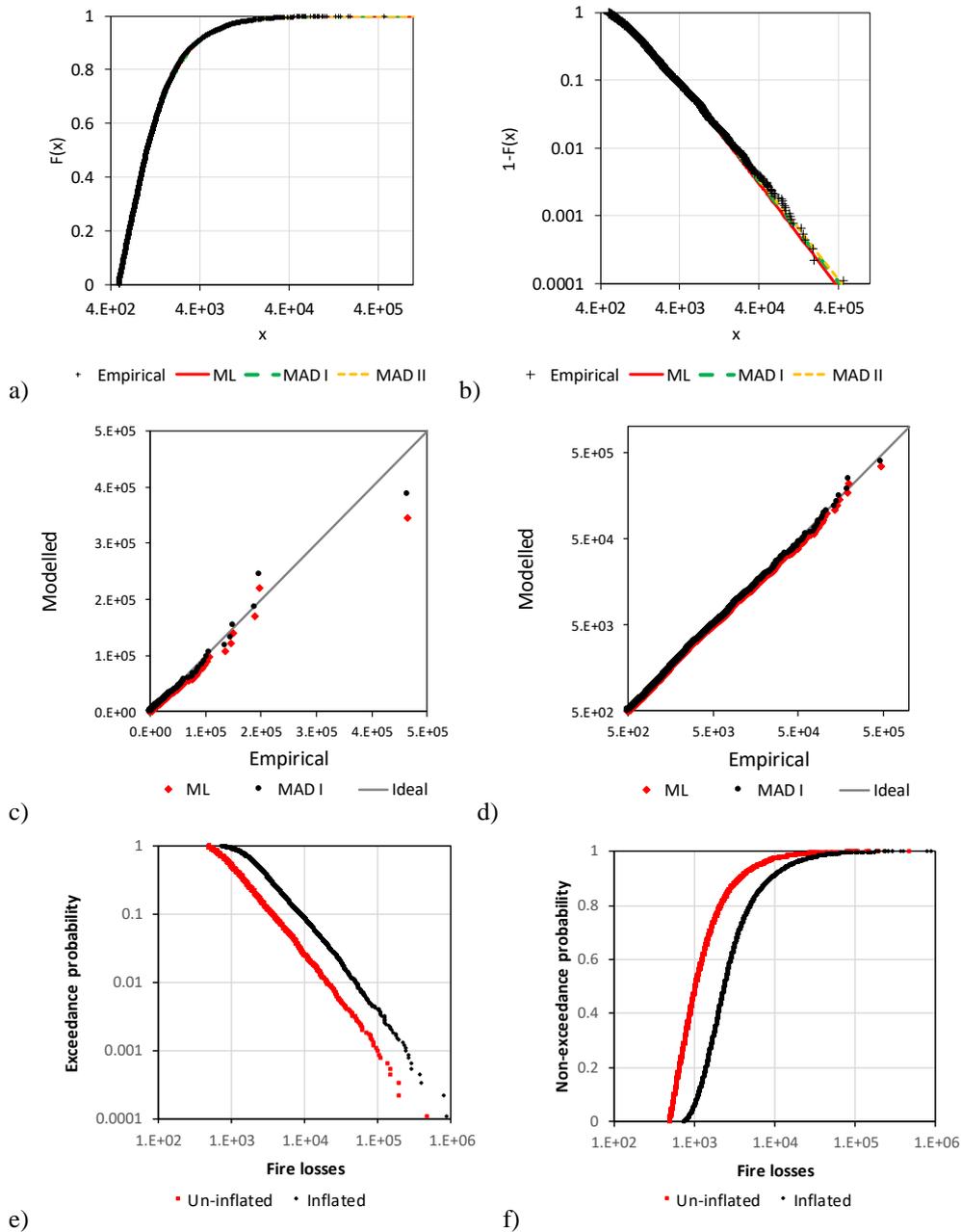

Figure 7: Analysis of Norwegian fire claim data: a) CDF of GPD, b) survival functions of GPD, c) Q-Q plot for the ML and MAD I estimation, d) Q-Q plot with logarithmic scale, e) empirical survival function for the original sample and the scaled sample (for 2012) and f) corresponding empirical CDF.

## *4.2 AON Re claim data*

Beirlant et al. (2004) applied a quantile regression to the claim data from AON Re Belgium ($n = 1823$), which are also provided by Dutang (2016). There is also a correlation $R = 0.329$ between the insured sum and the claim size in the entire data set and there is a cluster for smaller insured sums (Figure 8). Here, only claims for a sum insured ≥1E+7 are analysed with sample size $n$=1,069 being 59% of the original sample. This sub-sample does not include a significant correlation.

Firstly, the new tests of section 3.4 are applied. The highest 16 observations result in $m=12$ (30-31) with a corresponding Pareto-α of 1.251. The probability that a Pareto tail with this sample size has such an $m$ or higher is only 4.89%; an adjustment is reasonable.

A GPD is used as a base distribution and is fitted by the MAD method for the range between $x_{lower} = 3{,}500$ and $x_{upper} = 800{,}000$. The estimated parameters are $\hat{\gamma}_{base} = 1.792 \pm 0.099$ and $\hat{\sigma}_{base} = 1.122\text{E}+7 \pm 869.8$ The Weibull distribution (14) adopts the upper tail according to (5). The estimated transition probability is $\hat{P}_{upper} = 0.661 \pm 0.147$. The point estimation for the parameter of the adopting Weibull distributions are $\hat{\beta}_{adj,u} = 1.898$ and $\hat{\sigma}_{adj,u} = 9.144\text{E}+6$. The standard errors of these estimations are not informative as not all estimations of the 1000 resampling were stable. A share of 0.5% of the estimations included $\hat{P}_{upper}$ equal or very near to 0 and 2.8% of the estimations included $\hat{P}_{upper}$ equal or very near to 1. The corresponding values of $\beta_{adj,u}$ and $\sigma_{adj,u}$ are extreme. The estimation $\hat{\beta}_{adj,u}$ was restricted here on the range [0.5,100] and the estimated 90% confidence region is between 0.829 and 10.280. The 90% confidence region of $\sigma_{adj,u}$ is between 5.206E+6 and 3.879E+7. The bootstrap analysis shows that there is a certain probability that the actual unknown distribution is not adjusted with $\hat{P}_{upper} = 0$ or fully adjusted with $\hat{P}_{upper} = 1$. However, this does not exclude the point estimation in the sense of a significance test. The upper tail fits with the adaptation much better than without (s. Figure 9c and d).

The lower tail is adjusted by an additional GPD (13) with a negative extreme value index and a finite right endpoint being $x_{lower}$. In this way, only one additional parameter had to be estimated, which is the extreme value index of the adjusting GPD $\hat{\gamma}_{adj,l} = -0.750 \pm 0.106$. The scale parameter of the GPD is determined by $\hat{\sigma}_{adj,l} = x_{lower} \cdot \hat{\gamma}_{adj,l}$. The lower adjusted tail also fits the observations better than the unadjusted variant (s. Figure 9a and c).

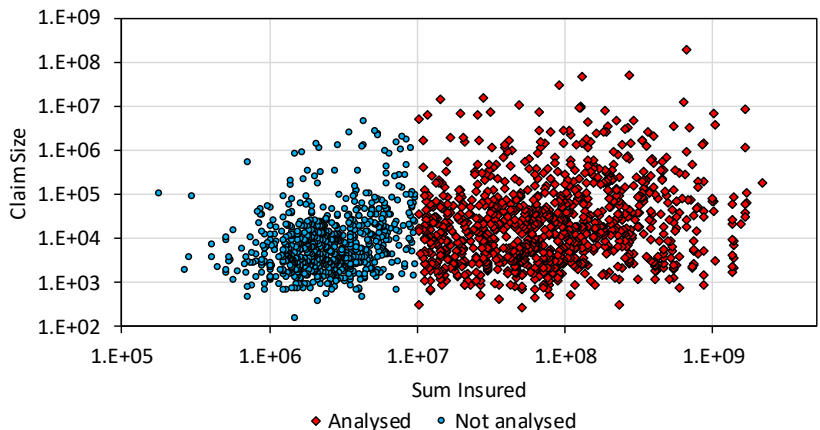

Figure 8: Claim size data of AON Re.

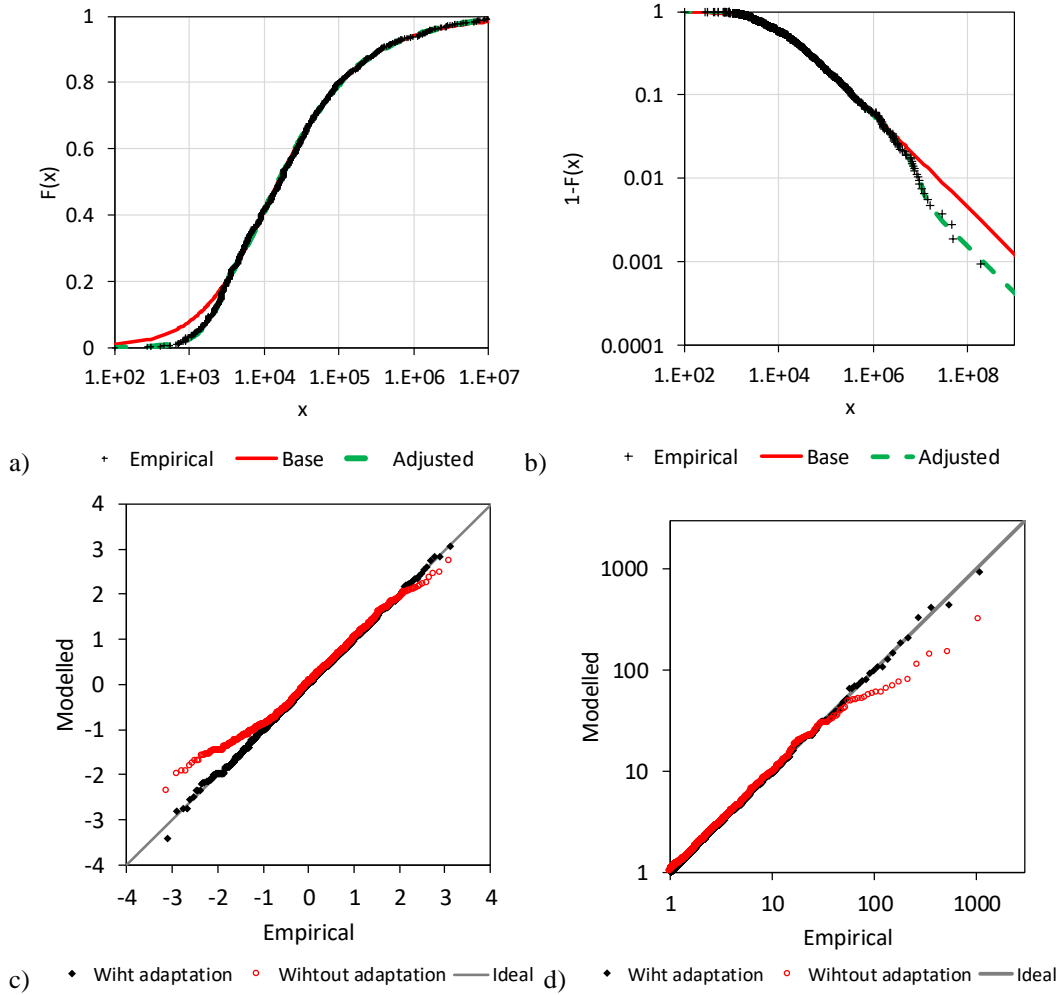

Figure 9: Analysis of AON Re Loss data: a) sum insured versus plot of claim size, b) CDF, c) Q-Q plot for standard normal margins and f) for standard Fréchet margins.

### 4.3 Danish fire loss data

In this section, the Danish fire loss sample of the R-package of Dutang (2016) is analysed. The smallest observation of the sample ($n = 2167$, millions DKK) is 1 million and is present 11 times in the sample. Therefore, the threshold of the base distribution is assumed at 999,999. The new test of section 3.4 results in a probability of 0.9% with $m \geq 23$, the observed value for the 27 highest observations with a corresponding Pareto-α of 1.524. Obviously, a simple Pareto tail is not likely.

The Pareto distribution is used as a base distribution and fitted by the MAD method for a range between $x_{lower} = 2.5\text{E}+6$ and $x_{upper} = 1.5\text{E}+7$. The estimation of the extreme value index is $\hat{\gamma}_{base} = 0.760 \pm 0.017$. A Weibull distribution (14) again realises the adjustment of the upper tail according to (5). The corresponding transition probability is $\hat{P}_{upper} = 0.404 \pm 0.184$. The point estimations for the parameter of the adapting Weibull distributions are $\hat{\beta}_{adj,u} = 4.842$ and $\hat{\sigma}_{adj,u} = 2.490\text{E}+7$. Once more, the standard errors of these estimations are not informative because not all estimations of the 1,000 resampling were stable. A share of 1.0% of the estimations included a probability equal or very near to 0, and 1.6% of the estimations included a probability equal or very near to 1. The corresponding values of $\beta_{adj,u}$ and $\sigma_{adj,u}$ are extreme. However, the bootstrap analysis shows again that there is a certain probability that the actual unknown distribution is not adjusted or fully adjusted, that does not exclude the point estimation in the sense of a significance test. The point estimation is just the 'middle' of extreme opportunities. The estimation of $\beta_{adj,u}$ was restricted as for the AON Re data, the 90% confidence region is between 1.688 and 48.837. The 90% confidence region of $\sigma_{adj,u}$ is between 1.705E+7 and 4.900E+7. The upper tail fits better with the adaptation ~~better~~ than without (s. Figure 10a).

As for the AON data, the lower tail is adjusted by an additional GPD (13) with a negative extreme value index and a finite right endpoint, being $x_{lower}$. The corresponding parameter is $\hat{\gamma}_{adj,l}$ =-0.435±0.118. The lower tail is also better fitted with the adjustment than without (Figure 10a). The Q-Q for standard normal margins transformed (Figure 10c) and standard Fréchet margins (Figure 10d) shows~~,~~ that the adaptation of the tail improves the fit. However, improvement of the lower tail is not as obvious as for the previous example. Also, the current data are not very good, the cluster of a claim size of 1 million distorts the analysis. But the influence is not very strong; this is an indication that the new approach is robust.

The Danish fire loss data has already been analysed in previous publications. However, Frigessi et al. (2002) and Bakar et al. (2015) did not present any plot of model and observations. Gian Paolo and Nino (2014) published different Q-Q plots. Their results do not seem to be better, their focus was on the performance of distance estimators, including the MAD method.

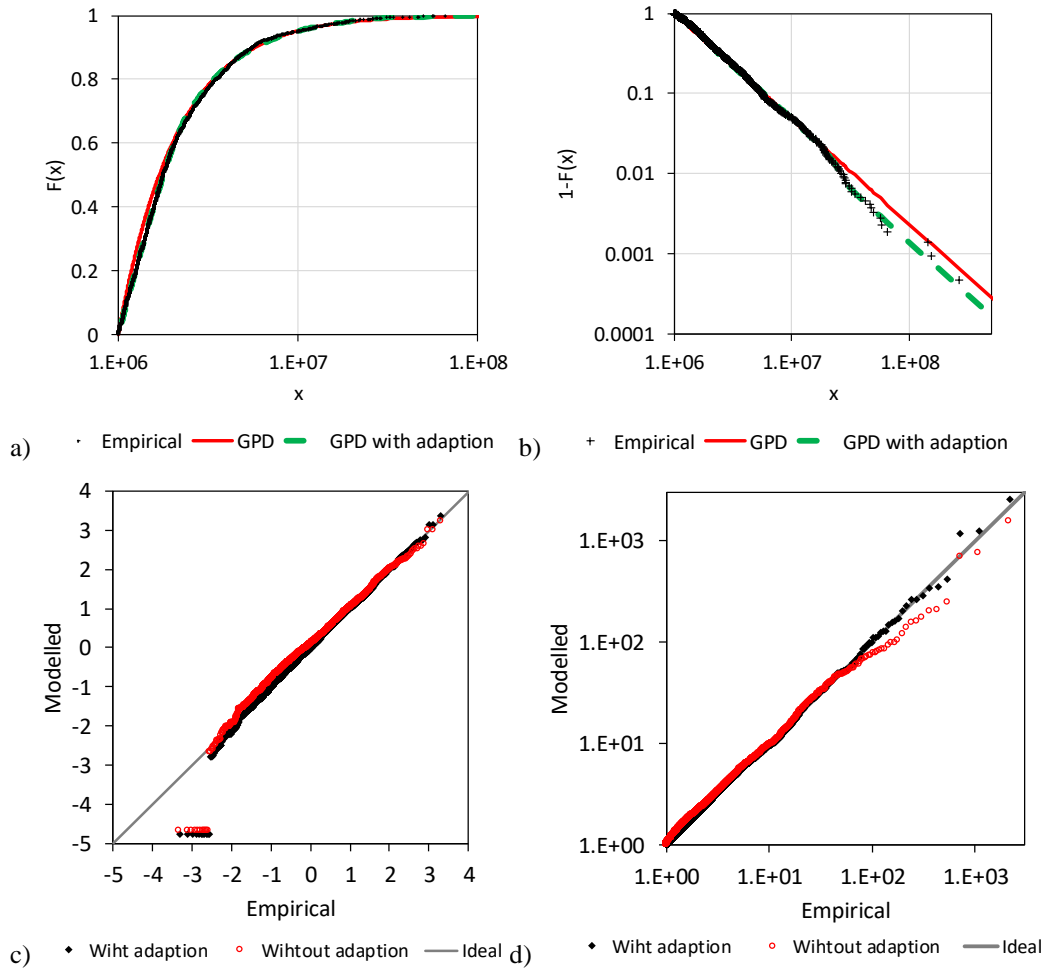

Figure 10: Analysis of the Danish fire loss data: a) CDF, b) survival function, c) Q-Q plot for standard normal margins and d) for standard Fréchet margins.

### 4.4 The Asia claim size data

In the last example, the tail of the commercial claim size data for Asia is analysed with 'usage' (occupancy class) 'Manufacturing' (Dutang 2016) and it is only briefly explained. The entire data are heterogenic according to Figure 11, different break points distort the distribution. The record threshold is probably not the same for all observations. Here, only the tail is modelled for all observations over the threshold of 2.2E+7 ($n = 51$). The test of the 14 largest observations with $m = 11$ results in a probability of 3.71%, with $m \geq 11$ for a Pareto distribution with a corresponding Pareto-α of 2.648; an adjustment is appropriate.

The Pareto distribution is used as base distribution and the Weibull distribution (14) for the adjustment of the upper tail (5). The MAD method is applied for the base distribution in range between $x_{lower} = 2.2E+7$ and $x_{upper} = 8.5E+7$. The analysis results are shown in Figure 12. The adjustment improves the fit considerably. However, the overall fit is not good since the data are probably not homogenous.

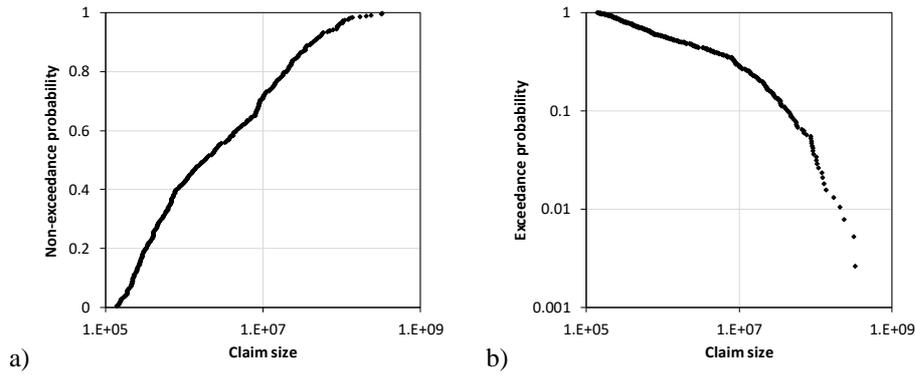

Figure 11: Empirical distribution of the Asian claim data (n=383, 'Manufacturing'): CDF, b) survival function.

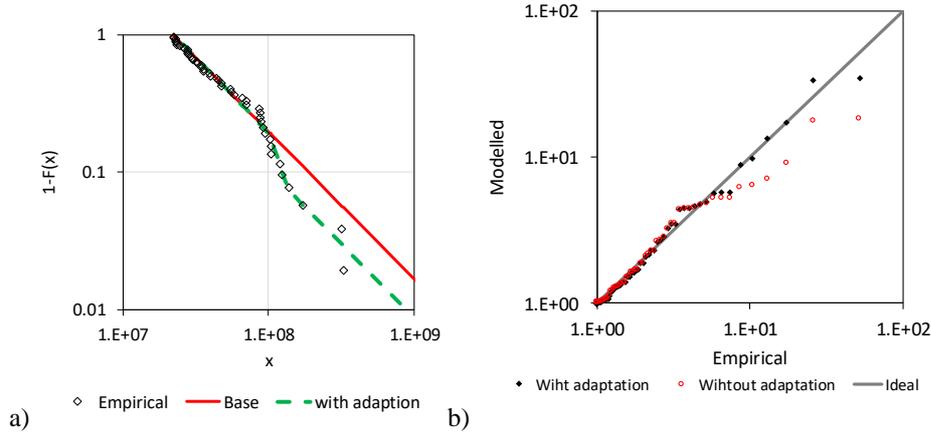

Figure 12: Analysis of the Asia claim size data: a) survival function for the tail ($\hat{\gamma}_{base}$ =0.953, $\hat{\beta}_{adj,u}$ =6.393, $\hat{\sigma}_{adj,u}$ =1.21E+8, $\hat{P}_{upper}$ =0.518), b) corresponding Q-Q plot with standard Fréchet margins.

## 5  Conclusion

In the current paper, the opportunities for modelling and estimating a claim size distribution have been extended. The motivation was the observation of extraordinary lower and upper tails of empirical claim or loss distributions. An adjustment of the upper and lower tails has been suggested in section 2, which implies an extension of the approaches of Meerschaert et al. (2012) and Raschke (2014) to lower tails and with the combination of discrete mixing (5). Claim generating stochastic mechanisms are formulated with conceivable interpretations for the claim management of an insurance policy in section 2.1. The Pareto distribution and GPD are assumed as base distributions that match well with actuarial practice (section 2.2). However, different stochastic mechanisms or approximations for the claim size distribution are possible according to section 2.3 and 2.5. This is bad news, since the claim generating process cannot be identified only by use of claim observations. It is also good news since the modelling is appropriate for different claim generating processes. The asymptotic behaviour of the adjusted upper tail was also discussed (section 2.4). The modelling is conservative since the final upper tail has the same Pareto-$\alpha$ as the base distribution if the transition probability $P_{upper}$ is smaller than 1. This is in contrast to the upper tail adjustments of Meerschaert et al. (2012) and Raschke (2014) which generate a higher Pareto-$\alpha$ for the very upper tail. (Raschke (2014) analysed exponential distributed earthquake magnitudes which correspond to Pareto distributed seismic moments, (cf. Beirlant et al. 2004). The differences between the very upper tail for $P_{upper} = 1$ and $P_{upper} < 1$ in Figure 2 and Figure 4 illustrate this.

The MAD method has been used for claim size distribution, as already done by Clemente et al. (2012), Skřivánková, and Juhás (2012) and Gian Paolo and Nino (2014) (section 3.1). One advantage of the MAD estimation is the opportunity of simple weighting. The normalised spacing was also explained in section 3 and extended the inference opportunity for Pareto-$\alpha$. The issue of goodness-of fit was discussed in section 3.4. Quantile plots have been suggested and applied and a new quantitative test for a Pareto tail is introduced to avoid over-fit.

The examples in section 4 have demonstrated that the extension of opportunities in modelling and inference are useful. The estimated confidence ranges of the examples emphasise that the point estimations are only opportunities in an extreme range between $P_{upper} = 0$ and $P_{upper} = 1$ for the upper tail adjustment. The issue: the behaviour of the very upper tail is completely and abruptly changed if $P_{upper}$ switch from <1 to 1. However, this could only concern probability ranges ~~being~~ outside of the interest of ~~an~~ actuarial analysis.

An important secondary result of the paper is that observed losses/claims do not need to be inflated for the estimation of a Pareto tail according to section 2.2. In contrast, un-inflated losses/claims can be used with a larger sample and a higher accuracy of the estimation of Pareto-$\alpha$ (cf. section 4.1).

For future research, the tail behaviour of such adjustments could be discussed in the community of extreme value theory and statistics. The inference methods could also be researched, and detail could be improved. The estimation error for the MAD method was estimated here by a boot strap analysis. It would be helpful~~,~~ if the estimation error (standard error) could be quantified without a Bootstrap analysis.